\documentclass[aps,prl,reprint,superscriptaddress]{revtex4-2}

\usepackage{graphicx}
\usepackage{amsmath}
\usepackage{amssymb}
\usepackage{bm}
\usepackage{xcolor}
\usepackage{hyperref}
\hypersetup{
  colorlinks=true,
  breaklinks=true,
  linkcolor=blue,
  urlcolor=blue,
  citecolor=blue,
}
\usepackage{changes}

\def\<#1>{\mathinner{\langle#1\rangle}}
\mathchardef\up="0222
\mathchardef\dn="0223
\DeclareMathOperator\imag{Im}

\begin{document}
\preprint{arXiv}

\title{Superconductivity and Mottness in Organic Charge Transfer Materials}

\author{Henri~Menke}
\affiliation{Max-Planck-Institut für Festkörperforschung, Heisenbergstraße 1, 70569 Stuttgart, Germany}
\affiliation{Department of Physics, University of Erlangen-N\"{u}rnberg, 91058 Erlangen, Germany}

\author{Marcel~Klett}
\affiliation{Max-Planck-Institut für Festkörperforschung, Heisenbergstraße 1, 70569 Stuttgart, Germany}

\author{Kazushi~Kanoda}
\affiliation{Max-Planck-Institut für Festkörperforschung, Heisenbergstraße 1, 70569 Stuttgart, Germany}
\affiliation{Physikalisches Institut, Universität Stuttgart, 70569 Stuttgart, Germany}
\affiliation{Department of Applied Physics, University of Tokyo, Hongo 7-3-1, Bunkyo-ku, Tokyo 113-8656, Japan}

\author{Antoine~Georges}
\affiliation{Coll\`ege de France, 11 place Marcelin Berthelot, 75005 Paris, France}
\affiliation{Center for Computational Quantum Physics, Flatiron Institute, New York 10010, USA}
\affiliation{CPHT, CNRS, Ecole Polytechnique, Institut Polytechnique de Paris, Route de Saclay, 91128 Palaiseau, France}
\affiliation{Department of Quantum Matter Physics, University of Geneva, 24 quai Ernest-Ansermet, 1211 Geneva, Switzerland}

\author{Michel~Ferrero}
\affiliation{CPHT, CNRS, Ecole Polytechnique, Institut Polytechnique de Paris, Route de Saclay, 91128 Palaiseau, France}
\affiliation{Coll\`ege de France, 11 place Marcelin Berthelot, 75005 Paris, France}

\author{Thomas~Sch{\"a}fer}
\email{t.schaefer@fkf.mpg.de}
\affiliation{Max-Planck-Institut für Festkörperforschung, Heisenbergstraße 1, 70569 Stuttgart, Germany}

\date{\today}

\begin{abstract}
The phase diagrams of quasi two-dimensional organic superconductors display a plethora of fundamental phenomena
associated with strong electron correlations, such as unconventional superconductivity, metal-insulator transitions,
frustrated magnetism and spin liquid behavior.
We analyze a minimal model for these compounds, the Hubbard model on an anisotropic triangular lattice, using cutting-edge quantum embedding methods respecting the lattice symmetry. We demonstrate the existence of unconventional superconductivity
by directly entering the symmetry-broken phase.
We show that the crossover from the Fermi liquid metal to the Mott insulator is associated with
the formation of a pseudogap.
The predicted momentum-selective destruction of the Fermi surface into
hot and cold regions provides motivation for further spectroscopic studies.
Our results are in remarkable agreement with experimental phase diagrams of $\kappa$-BEDT organics.
\end{abstract}

\maketitle

\textit{Introduction.}
Strongly correlated materials in which mutual interactions between electrons drive the physics exhibit some of the most fascinating collective phenomena of condensed matter physics. Their phase diagrams is extremely rich, hosting unconventional superconductivity, quantum criticality, and quantum magnetism. This plethora of phases hints at the many competing energy scales at play, which make these materials highly sensitive to small parameter changes, such as strain, pressure, chemical substitutions and doping, also an appealing feature for potential technological applications.

The organic charge-transfer salts $\kappa$-(BEDT-TTF)$_2$X and $\kappa$-(BETS)$_2$X form a very interesting family of such materials -- see \cite{Kanoda2011,Powell2006,Powell2011,Riedl2022} for reviews.
Here BEDT-TTF and BETS are, respectively, the organic electron donor molecules C$_{10}$H$_8$S$_8$ and C$_{10}$H$_8$S$_4$Se$_4$
and X is an inorganic electron acceptor molecule \cite{Riedl2022,Powell2006,Powell2011}. The phase diagram of these compounds can be exquisitely controlled by either varying pressure
or changing the acceptor X.
It hosts paradigmatic examples of many-body physics such as Mott metal-insulator transitions \cite{Lefebvre2000,Limelette2003,Kagawa2004,Shimizu2007,Faltermeier2007,Merino2008,Dumm2009,Ivek2017,Saito2021,Pustogow2021}, and associated (quantum) criticality \cite{Kagawa2009,Terletska2011,Vucicevic2013,Vucicevic2015,Furukawa2015,Wakamatsu2023},
putative quantum spin liquid regimes \cite{Powell2011,Shimizu2003,Saito2021,Pustogow2022,Wakamatsu2023}, magnetic phases \cite{Miyagawa1995,Lefebvre2000}, and unconventional superconductivity \cite{Wosnitza2019,Powell2006,Powell2011,Lefebvre2000,Shimizu2010,Cavanagh2019,Imajo2021}.

From a theoretical point of view the dimerization of the organic molecules opens the possibility of a simplified
description in terms of an  effective half-filled single-band Hubbard model \cite{Hubbard1963,Hubbard1964,Gutzwiller1963,Kanamori1963,Arovas2022,Qin2022}
on an anisotropic triangular lattice
\cite{Powell2006,Powell2011,Zantout2018,Kyung2006,Hebert2015,Acheche2016,Yu2023} --
although it has been suggested that some properties require a two-band generalization of this
model~\cite{Guterding2015,Riedl2022}.

Due to the competing instabilities present in this model and its low dimensionality,
both on-site temporal quantum fluctuations and spatial correlations have to be properly taken into account. Cluster extensions \cite{Maier2005,Kotliar2006,Tremblay2006} of the dynamical mean-field theory (DMFT, \cite{Georges1996}) are able to treat both types of fluctuations, the temporal ones exactly, the spatial ones up to distances set by the cluster size $N_c$. These techniques have proven particularly useful in the description of strongly correlated superconductivity and metal-insulator transitions in magnetically frustrated systems \cite{Parcollet2004,Kyung2006b,Hebert2015,Fratino2016,Downey2023,Downey2023b,Meixner2023,Tscheppe2023}.

\begin{figure*}[t!]
  \centering
  \includegraphics[width=\linewidth]{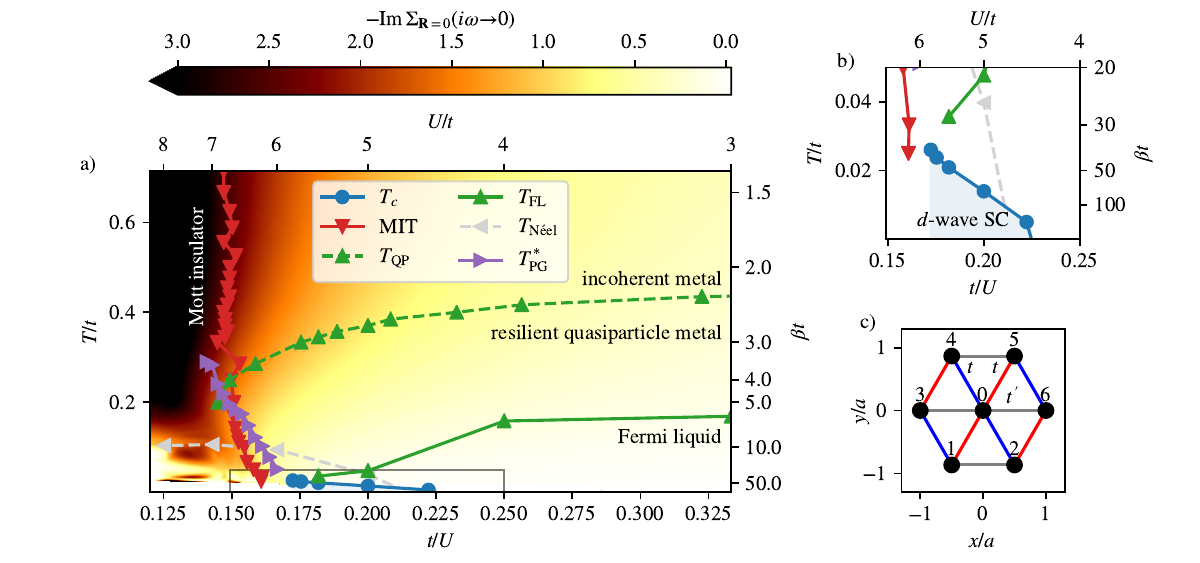}
  \caption{a) Phase diagram of the half-filled anisotropic triangular lattice Hubbard model with $t'=0.4t$ calculated in CDMFT with $N_c=7$, as a function of temperature $T/t$ and $t/U$.  b) Zoom on the superconducting regime. c) Depiction of the $N_c=7$-site cluster.
  }
  \label{fig:phasediagram}
\end{figure*}

In this article we establish the phase diagram and the spectral properties of the anisotropic triangular lattice Hubbard model by cutting-edge cellular DMFT (CDMFT) calculations with center-focused post processing \cite{Klett2020}. We use cluster sizes $N_c=7$ that go beyond the usual $N_c=4$ plaquette and respect the triangular lattice symmetry [see Fig.~\ref{fig:phasediagram}c)]. With our calculations we track the continuous destruction of the metallic state with increasing interaction strengths and trace it back to the development of hot regions of enhanced quasiparticle scattering on the Fermi surface. By entering the U(1)-symmetry-broken phase we demonstrate that a superconducting phase with d-wave symmetry is established in the vicinity of the Mott metal-insulator crossover. The superconducting phase emerges out of a Fermi liquid phase, except very close to the metal-insulator
transition where a pseudogap emerges which competes with superconductivity. At the end of our paper we show that the superconducting transition temperature and further phase diagram features obtained are in remarkable agreement with experiments on organic charge-transfer salts.

\textit{Model and method.}
We consider the single-band Hubbard model on the triangular lattice
\begin{equation}
  H
  = - \sum_{\<i,j>,\sigma} (t_{ij} c_{i,\sigma}^\dagger c_{j,\sigma} + \mathrm{h.c.})
  - \mu \sum_{i,\sigma} n_{i,\sigma}
  + U \sum_{i} n_{i,\up} n_{i,\dn},
\end{equation}
where $\sigma$ is the electronic spin degree of freedom, $c_{i,\sigma}$ is the creation operator of an electron on site $i$ with spin $\sigma$, and $n_{i,\sigma} = c_{i,\sigma}^\dagger c_{i,\sigma}$ is the number operator.

In our model calculations we choose $t_{ij} = t'$ on the horizontal bonds and $t_{ij}=t$ otherwise (see Fig.~\ref{fig:phasediagram}c)].  The limit $t = t'$ corresponds to the fully frustrated triangular lattice, whereas $t \neq t'$ introduces an anisotropy that breaks the lattice symmetry, and $t'=0$ represents a square lattice rotated by $\pi/4$ as a limiting case. The chemical potential is given be $\mu$ and $U$ quantifies the magnitude of the local Coulomb repulsion.  Further we choose $t$ as our unit of energy.

We analyze the model by means of (paramagnetically restricted) cellular dynamical mean-field theory, partially with center-focused post processing \cite{Klett2020,Meixner2023} with $N_c = 7$ sites, i.e., a central site and one shell of nearest neighbors, cf.~Fig.~\ref{fig:phasediagram}c). Further computational details can be found in the Supplemental Material \cite{Supplemental}.

\textit{Phase diagram.}
\begin{figure*}[t!]
  \centering
  \includegraphics[width=\linewidth]{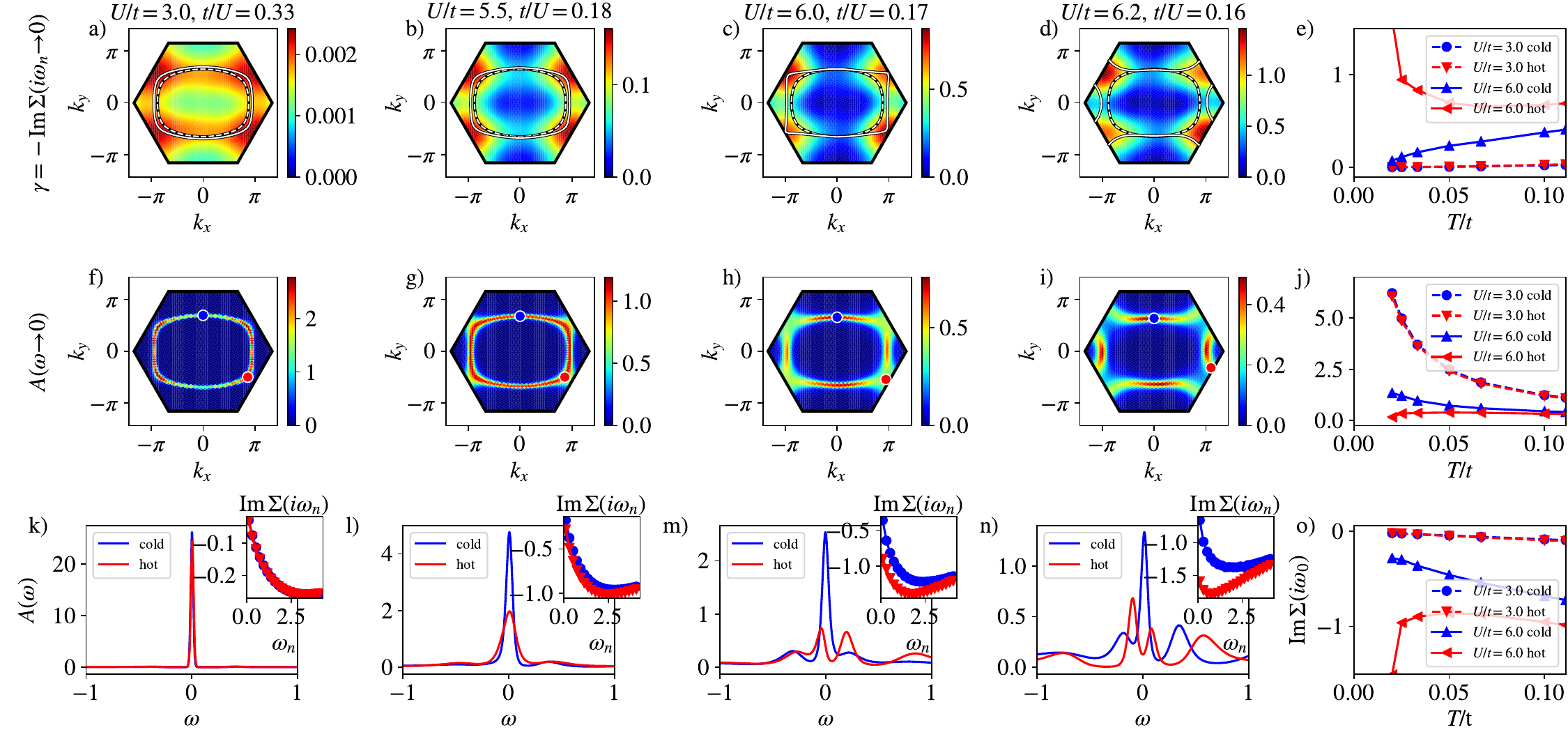}
  \caption{Spectral properties for $t'=0.4t$. a)--d) Quasiparticle scattering rate $\gamma$ for increasing values of the interaction from left to right at $T/t=0.033$. The solid (dashed) line corresponds to the (non-)interacting Fermi surface. e) $\gamma$ at the hot and cold spot on the Fermi surface as a function of temperature.  f)--i) Spectral weight at the Fermi level for increasing values of the interaction from left to right at $T/t=0.033$. Blue (red) dots indicate the cold (hot) spots. j) Spectral weight at the Fermi level at the hot and cold spot as a function of temperature.  k)--n) Analytical continuation of the Green function in self-energy periodization at the hot and cold spot.  The insets show the periodized self-energy at the first few Matsubara frequencies.  o) ``First Matsubara frequency rule'' \cite{Chubukov2012,Maslov2012} at the hot and cold spot as a function of temperature.}
  \label{fig:periodization}
\end{figure*}
We explore the phase diagram as a function of temperature $T/t$ and the inverse ratio of hopping amplitude and interaction strengths $t/U$, corresponding to the application of pressure to the system.
Fig.~\ref{fig:phasediagram} shows the phase diagram for half-filling and $t'=0.4t$, approximately corresponding to realistic modellizations of $\kappa$-(BEDT-TTF)$_2$Cu[N(CN)$_2$]Cl ($\kappa$-Cl) and $\kappa$-(BEDT-TTF)$_2$Cu[N(CN)$_2$]Br ($\kappa$-Br) \cite{Kandpal2009,Nakamura2009}. For both materials $t \approx 67$ meV \cite{Guterding2015}. In the Supplemental Material \cite{Supplemental} we show the analogous phase diagram and data for $t'=0.1t$, a less frustrated system.

Let us first lay out the general arrangement of the different phases and regimes, before we discuss in detail how these have actually been determined. At small values of the interaction and high temperatures we find an incoherent regime which upon cooling below $T_\text{QP}$ evolves into a metal with `resilient' quasiparticles \cite{Deng2013} and, at
a lower temperature $T<T_\text{FL}$, a Fermi liquid.
Upon decreasing pressure, i.e., the relative bandwidth $t/U$, 
at intermediate temperatures, the metallic state is continuously renormalized until a pseudogap opens for $T<T^*_\text{PG}$, the precursor of a Mott metal-insulator crossover (MIT). At the ``foothills" of the metal-insulator crossover line, on its metallic side and for $T<T_c$, we observe a stable $d_{xy}$ superconducting phase, with a maximum $T_c/t \approx 0.024$. This calculated phase diagram is indeed in remarkable qualitative agreement to the experimentally determined ones of $\kappa$-Cl \cite{Riedl2022} and $\kappa$-Br \cite{Furukawa2023}.

\textit{Metallic and Fermi liquid regimes at larger bandwidth.}
We investigate the one-particle spectral properties at representative values of $t/U$.
At large $t/U=0.33$ quasiparticle coherence is established upon cooling the system below $T \leq T_\text{QP} \approx 0.45t$, which is indicated by a sign change of the slope of the imaginary part of the self-energy of the center site of the cluster at $\imag\Sigma_{\mathbf{R}=0}(i\omega_n)$ for $\omega_n \to 0$ \cite{Schaefer2015,Schaefer2016b,Simkovic2020,Schaefer2021}. Although `resilient' quasiparticles \cite{Deng2013} already exist in this regime, true Fermi-liquid behavior only sets in at a lower $T \leq T_\text{FL} \approx 0.15t$.

For the determination of $T_\text{FL}$ we switch to momentum space via the periodization of the self-energy, see \cite{Supplemental}. Fig.~\ref{fig:periodization}a) shows the quasiparticle scattering rate $\gamma_{\mathbf{k}}=-\imag \Sigma(\mathbf{k},i\omega_n \to 0)$ in the first Brillouin zone, obtained as a result of this periodization, at $T/t=0.033$.
One can see that for $t/U=0.33$ the momentum-dependence of $\gamma_{\mathbf{k}}$ is relatively moderate, both over the entire Brillouin zone as well as over the Fermi surface (indicated by a white solid line). Fig.~\ref{fig:periodization}f) displays the corresponding spectral weight at the Fermi level, $A_{\mathbf{k}}(\omega \to 0)=-\frac{1}{\pi} \imag G(\mathbf{k},i\omega_n \to  0)$, obtained by a linear extrapolation of the first two Matsubara frequencies of the Green function. The spectral weight $A_{\mathbf{k}}$ is only moderately varying over the Fermi surface. The maxima (minima) of the scattering rate $\gamma_{\mathbf{k}}$ on the correlated Fermi surface define the hot (cold) spots $\mathbf{k}_\text{hot}$ ($\mathbf{k}_\text{cold}$), indicated by red (blue) dots in the figure. For both of these respective Fermi surface points, at $t/U=0.33$, we observe that the scattering rate drops when cooling the system [Fig.~\ref{fig:periodization}e)], whereas the spectral weight is concomitantly increasing [Fig.~\ref{fig:periodization}j)], corroborating the notion of a metallic regime.

In order to decide whether this metal is a Fermi-liquid, we employ the ``first Matsubara frequency rule''~\cite{Chubukov2012,Maslov2012}, i.e., we plot $\imag\Sigma(\mathbf{k},i\omega_0)$ as a function of temperature [Fig.~\ref{fig:periodization}o)], and determine $T_\text{FL}$ as the temperature below which this quantity depends linearly on $T$ for all points on the Fermi surface (especially $\mathbf{k}_\text{hot}$). As stated before, for $t/U=0.33$ this is the case at $T \leq T_\text{FL} \approx 0.15t$,
substantiated by the one-particle spectra on the real axis [Fig.~\ref{fig:periodization}k)], obtained by MaxEnt analytic continuation \cite{Kraberger2017}. These display a clear quasiparticle peak for both, hot and cold spot. The inset shows that also the the imaginary part of the self-energy on the Matsubara axis displays metallic behavior.

\textit{Pseudogap and metal-insulator crossover.}
Let us now analyze the impact of increased interaction strength, shown in the remaining panels of Fig.~\ref{fig:periodization}. We can immediately observe an increased value of the quasiparticle scattering rate $\gamma_{\mathbf{k}}$ for stronger interactions, as well as a much more pronounced momentum-differentiation [Fig.~\ref{fig:periodization}b)-d)]. The effect is most visible considering the difference in $\gamma_{\mathbf{k}}$ at the hot and cold spots on the Fermi surface, and immediately leads, in turn, to an increased momentum-differentiation of the spectral weight for intermediate $t/U=0.18$ and $0.17$ [Fig.~\ref{fig:periodization}b)-c)]. For $t/U=0.16$ also the temperature-dependence of the quasiparticle scattering rate is drastically modified [Fig.~\ref{fig:periodization}e)]: whereas at the cold spot $\gamma_{\mathbf{k}}$ decreases with decreasing temperature, at the hot spot we can observe an enhancement for $T/t \lesssim 0.05$.

This enhancement of $\gamma_{\mathbf{k}_\text{hot}}$ has the immediate consequence that the spectral weight $A_{\mathbf{k}}$ is suppressed at the hot spot but not at the cold spot, which is the single-particle manifestation of  the emergence of a pseudogap. Our finding is indeed corroborated by the analytically continued spectral function [Fig.~\ref{fig:periodization}m)], which shows a clear quasiparticle peak for the cold spot, and a reduced, more incoherent spectrum at the hot spot.

\begin{figure}[t!]
    \includegraphics[width=\linewidth]{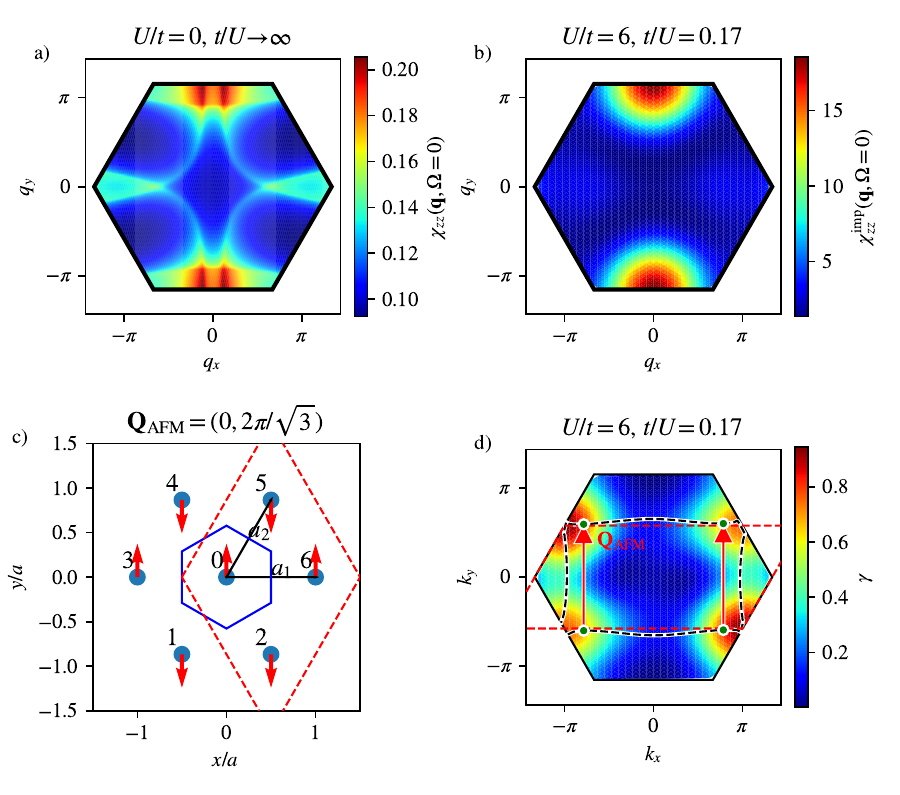}
    \caption{Magnetism and nesting for $t'=0.4t$ at $T/t=0.025$. a) Static (Lindhard) susceptibility of the non-interacting model. b) Static impurity susceptibility. c) Stripy antiferromagnetic ordering with Wigner-Seitz and magnetic unit cell in real space. d) Correlated Fermi surface with points nested by the antiferromagnetic ordering wave vector in green and Wigner-Seitz and magnetic unit cell in reciprocal space.}
    \label{fig:nesting}
\end{figure}

At the largest interaction value shown in Fig.~\ref{fig:periodization}, $t/U=0.16$, the scattering processes between the quasiparticles are already so strong, that the Fermi surface eventually becomes topologically reconstructed from its original electron-like shape to hole-like (Lifshitz transition) [Fig.~\ref{fig:periodization}i)]. Here the self-energy is very large for low Matsubara frequencies at the hot spot [inset of Fig.~\ref{fig:periodization}i)].

For larger values of the interaction a gap is developing for all Fermi surface points, culminating in the emergence of a Mott-insulating phase. At all temperatures that we could reach, this metal-insulator gapping happens as a crossover, rather than a thermodynamic phase transition (in contrast to CDMFT on smaller clusters \cite{Hebert2015}).
 The crossover line as a function of interaction and temperature is mapped out in Fig.~\ref{fig:phasediagram}, marked by the inflection point of the spectral weight at the Fermi level at the central site $A_{\mathbf{R}=0}(\omega \to 0)$ as a function of $U$ for fixed $T$. In this figure we also show the local scattering rate $\gamma_{\mathbf{R}=0}=-\imag\Sigma_{\mathbf{R}=0}(i\omega_n \to 0)$ as a background shading.

\textit{Magnetism and spin dynamics.} An interesting question is whether the location of the hot regions is a consequence of Fermi surface nesting, despite their occurrence at moderate to strong coupling strengths. Fig.~\ref{fig:nesting}a) shows the non-interacting static spin-spin susceptibility $\chi_{zz}(\mathbf{q},i\Omega_n=0)$ (Lindhard bubble) for $t'=0.4t$ and $T=0.025t$. We can see that, for this non-interacting situation,  $\chi_{zz}$ is (incommensurably) peaked around $\mathbf{q}=\mathbf{M}=(\pi,0)$, a vector corresponding to stripy antiferromagnetism with ferromagnetic correlations along the bonds with `suppressed' hopping $t'$ [Fig.~\ref{fig:nesting}c)]. We also show the interacting susceptibility in momentum space $\chi_{zz}(\mathbf{q},i\Omega_n=0)$ at $t/U=0.17$ [Fig.~\ref{fig:nesting}b)] obtained from the cluster susceptibility by periodization \cite{Supplemental}. $\chi_{zz}$ is peaked at the (commensurate) $\mathbf{M}$-point \footnote{For resolving finer details (like the precise incommensurability) larger cluster sizes or a Bethe-Salpeter treatment would have to be considered.}.

As can be seen from Fig.~\ref{fig:nesting}d), this $\mathbf{M}$-point vector effectively nests significant horizontal
portions of the Fermi surface. The hot regions are however confined to areas near the BZ boundaries. We attribute this phenomenon to the dual role of these areas: they are interconnected via the nesting vector and simultaneously situated in proximity to a van Hove singularity, characterized by a flattened energy landscape, thereby expanding the scattering phase space.

Also, let us comment here that, if we had not restricted our calculations to the SU(2)-symmetric solution, magnetic order with $\mathbf{q}=\mathbf{M}$ would have occurred, indicated by the dashed grey $T_\text{N{\'e}el}$-line in Fig.~\ref{fig:phasediagram}. More calculations of the (non-interacting bubble) susceptibility for different values of $t'/t$ can be found in the Supplemental Material \cite{Supplemental}. There, in the phase diagram for $t'=0.1t$ one can see that a reduction of the frustration parameter $t'/t$ leads to an inflation of the magnetically ordered regime in the interacting model. Valuable additional information of the spin dynamics can be obtained from nuclear magnetic resonance (NMR), discussed in the Supplemental Material \cite{Supplemental}.

\textit{Unconventional superconductivity.}
We now turn to the analysis of the superconducting dome at the ``foothills" of the Mott metal-insulator crossover at intermediate-to-high interaction strengths. At elevated temperatures we first search for the response of the system to different symmetries ($s$, $p$, $d_{xy}$, $d_{x^2-y^2}$) of an applied pairing field~\cite{Koretsune2005}. The strongest response we obtained from a field with $d_{xy}$ symmetry, whereas the systems is barely susceptible to other symmetries \footnote{The $d_{xy}$ symmetry on the triangular lattice corresponds to $d_{x^2-y^2}$ symmetry on the square lattice by coordinate transformation. Please note that, particularly in the theoretical literature, superconducting order parameters are often defined in coordinated systems rotated by $\pi/4$ from the crystal axes. In this basis, the $d_{xy}$ and $d_{x^2-y^2}$ labels are reversed \cite{Cavanagh2019}.}. For determining the exact nature of the superconducting regime we enter the symmetry-broken phase directly and measure the anomalous Green function $F$ (see \cite{Supplemental} for respective plots).
We further map out the shape of the superconducting transition, i.e., the dependence of $T_c$ on $t/U$ (see Fig.~\ref{fig:phasediagram}): a finite $T_c$ is obtained for $t/U \lesssim 0.22$ until the Mott crossover sets in. For the parameters analyzed we obtain a maximum $T_c^\text{max}/t \approx 0.024$ at $t/U \approx 0.17$. This, quite remarkably, places the superconducting phase right in the vicinity of the Mott crossover.

The mechanism of spin-fluctuation mediated superconductivity in the context of layered organic superconductors has been discussed early on in \cite{Schmalian1998}: in the vicinity of the Mott crossover (where strong quasiparticle scattering is present, cf. Fig.~\ref{fig:periodization}) spin fluctuations are enhanced by increased coupling strength below $T_\text{N{\'e}el}$, being able to act as a pairing glue. Let us stress here that, despite the close vicinity to the Mott crossover, the superconducting phase is always entered from a (renormalized) Fermi liquid. 
Approaching the Mott crossover, the spectral weight of quasiparticles is strongly suppressed by the opening of the 
pseudogap. This is expected to suppress superconductivity, and the  competition between superconductivity and the pseudogap was 
previously reported for the square lattice using the dynamical cluster approximation \cite{Gull2013} and indeed we 
report in \cite{Supplemental} a superconducting dome for $t'/t=0.1$. For $t'/t=0.4$ this drop of $T_c$ likely happens 
in a very narrow range near the Mott insulator that we have not resolved.  This effect is, in fact, very reminiscent of the strongly correlated mechanism of superconductivity in fullerides \cite{Capone2002,Capone2009}, although, there, Jahn-Teller phonons are an additional crucial ingredient.

\textit{Relation to experiments.}
In order to compare to experimental results we note that the transfer integrals $t$ for the dimerized $\kappa$-Cl 
and $\kappa$-Br are roughly equal, with $t \approx 67\,\mathrm{meV}$ \cite{Guterding2015}. We also note that, according to \cite{Limelette2003}, increasing pressure by 300bar=30MPa corresponds to a 3 percent increase of $t/U$.
The phase diagram of the two compounds are qualitatively similar, exhibiting superconducting as well as antiferromagnetically ordered regimes close to a Mott transition \cite{Riedl2022,Furukawa2023}. More quantitatively, the highest superconducting transition temperature is $T_c=13.4\,\mathrm{K}$ ($\kappa$-Cl \cite{Riedl2022}) and $T_c=11.6\,\mathrm{K}$ ($\kappa$-Br \cite{Furukawa2023}), which compares excellently with our computational result of $T=18\,\mathrm{K}$.

For magnetic and Fermi surface properties one has to keep in mind that the $\kappa$-type molecular arrangement can be modeled into an isosceles triangular lattice as depicted by Fig.~\ref{fig:phasediagram}c); however, it contains two differently oriented dimers, which sit alternatively at the lattice points. Therefore, the unit cell of $\kappa$-(BEDT-TTF)$_2$X is doubled and the first Brillouin zone is folded into a rectangle. The $^{13}$C NMR relaxation rate, which probes the intensity of antiferromagnetic fluctuations, increases when the system approaches the Mott-transition critical pressure from the metallic side by decreasing pressure \cite{Furukawa2023}. 

This feature is well reproduced by the present calculations shown in the Supplemental Material \cite{Supplemental}. It is expected that the enhanced fluctuations occur at the wave vector 
associated with the long-range antiferromagnetic order of the Mott insulator, i.e., stripy-spin magnetism as shown in Fig.~\ref{fig:nesting}c) \cite{Lefebvre2000,Miyagawa2002,Smith2004,Yusuf2007,Kagawa2009,Ishikawa2018}. The regions connected by this wave vector on the Fermi surface are magnetically activated spots and are well in accord with the locations of the predicted hot spots shown in Fig.~\ref{fig:periodization}b)-d) \cite{Furukawa2023}. For further experimental determination of the location of hot and cold spots studies with angle-resolved photoemission spectroscopy (ARPES), quantum oscillations, and angular-dependent magnetoresistance (ADMR) are highly desirable.

\textit{Conclusions and outlook.}
We performed cutting-edge quantum-embedding calculations for models of $\kappa$-organic compounds. We found excellent agreement of our computational results with the experimental phase diagrams of $\kappa$-Cl and $\kappa$-Br. Our work paves the way to future theoretical studies on quantum spin-liquid candidate systems such as $\kappa$-(BEDT-TTF)$_2$Cu$_2$(CN)$_3$ \cite{Pustogow2023} with a higher degree of frustration $t'/t$.

\begin{acknowledgements}
\textit{Acknowledgements.}
We acknowledge fruitful discussions with Sabine Andergassen, Martin Dressel, Philipp Hansmann, Elio K{\"o}nig, Andrew J. Millis, Andrej Pustogow, Andr{\'e}-Marie Tremblay, Roser Valent{\'i},
and Nils Wentzell.
The authors gratefully acknowledge the scientific support and HPC resources provided by the Erlangen National High Performance Computing Center (NHR@FAU) of the Friedrich-Alexander-Universität Erlangen-Nürnberg (FAU)
(hardware funded by the German Research Foundation - DFG),
the computing service facility of the MPI-FKF and the
Scientific Computing Core of the Flatiron Institute. The Flatiron Institute is a division of the Simons Foundation.
\end{acknowledgements}

\bibliographystyle{unsrturl}
\bibliography{references.bib}

\end{document}


\title{Superconductivity and Mottness in Organic Charge Transfer Materials\\
\textit{-- Supplemental Material --}}

\author{Henri~Menke}
\affiliation{Max-Planck-Institut für Festkörperforschung, Heisenbergstraße 1, 70569 Stuttgart, Germany}
\affiliation{Department of Physics, University of Erlangen-N\"{u}rnberg, 91058 Erlangen, Germany}

\author{Marcel~Klett}
\affiliation{Max-Planck-Institut für Festkörperforschung, Heisenbergstraße 1, 70569 Stuttgart, Germany}

\author{Kazushi~Kanoda}
\affiliation{Max-Planck-Institut für Festkörperforschung, Heisenbergstraße 1, 70569 Stuttgart, Germany}
\affiliation{Physikalisches Institut, Universität Stuttgart, 70569 Stuttgart, Germany}
\affiliation{Department of Applied Physics, University of Tokyo, Hongo 7-3-1, Bunkyo-ku, Tokyo 113-8656, Japan}

\author{Antoine~Georges}
\affiliation{Coll\`ege de France, 11 place Marcelin Berthelot, 75005 Paris, France}
\affiliation{Center for Computational Quantum Physics, Flatiron Institute, New York 10010, USA}
\affiliation{CPHT, CNRS, Ecole Polytechnique, Institut Polytechnique de Paris, Route de Saclay, 91128 Palaiseau, France}
\affiliation{Department of Quantum Matter Physics, University of Geneva, 24 quai Ernest-Ansermet, 1211 Geneva, Switzerland}

\author{Michel~Ferrero}
\affiliation{CPHT, CNRS, Ecole Polytechnique, Institut Polytechnique de Paris, Route de Saclay, 91128 Palaiseau, France}
\affiliation{Coll\`ege de France, 11 place Marcelin Berthelot, 75005 Paris, France}

\author{Thomas~Sch{\"a}fer}
\email{t.schaefer@fkf.mpg.de}
\affiliation{Max-Planck-Institut für Festkörperforschung, Heisenbergstraße 1, 70569 Stuttgart, Germany}

\maketitle

In this Supplemental Material we first give details on how the computations in the main text have been performed (Sec.~\ref{sec:computation}), then we show additional results for magnetism and spin dynamics (Sec.~\ref{sec:magnetism}) before we eventually report results of calculations at lower frustration $t'=0.1t$ (Sec.~\ref{sec:additional}).

\section{Computational details}
\label{sec:computation}

\subsection{Geometry of the cluster}

\begin{figure}[tb]
  \centering
  \unitlength=20.5pc
  \begin{picture}(1,.53)
    \put(0,0){\includegraphics[width=.5\unitlength]{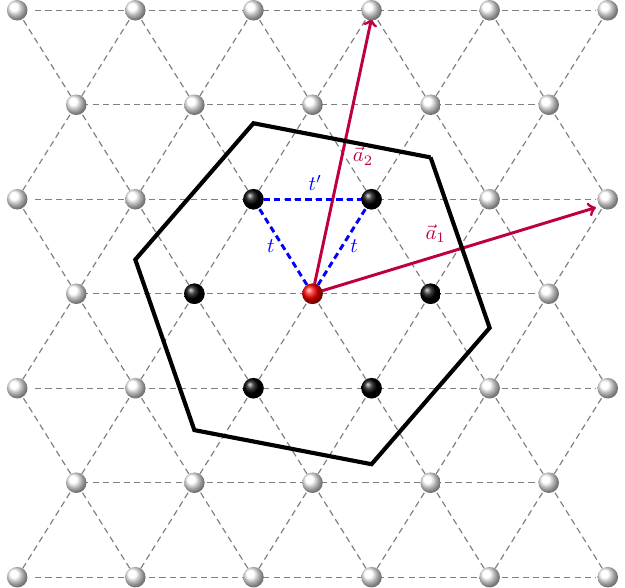}}
    \put(0,.5){a)}
    \put(.5,0){\includegraphics[width=.5\unitlength]{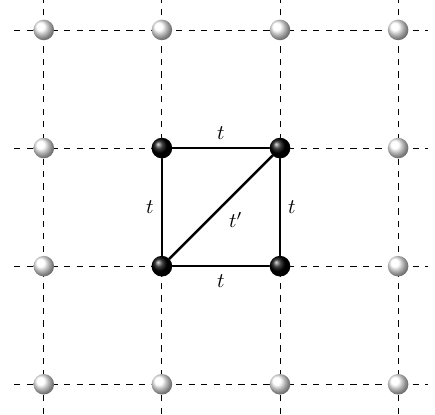}}
    \put(.5,.5){b)}
  \end{picture}
  \caption{(a)~Seven site cluster with full $C_6$ rotational symmetry around the center. (b)~Triangular lattice represented by a square lattice geometry with even number of sites and hopping among a single diagonal direction.}
  \label{fig:cluster}
\end{figure}

In performing our cellular DMFT calculations two different geometries of the cluster are chosen, both of which are shown in Fig.~\ref{fig:cluster}. On the one hand, for most of the calculations presented, we use a cluster with seven sites that comprises a central site and one ``shell'' of nearest neighbors around it, thereby preserving the full $C_6$ rotational symmetry of the lattice around the central site, cf.~Fig.~\ref{fig:cluster}~a).  As has been shown in Ref.~\cite{Klett2020} the physical properties of the central site in the cluster will converge faster towards the value of the lattice problem upon increasing the cluster size.  Therefore, for the estimation of single-particle observables we will only focus on the central site in the seven site cluster.

However, this seven site cluster has an important shortcoming in the context of magnetic instabilities. Any antiferromagnetic order is incommensurate with a cluster containing an odd number of sites and will therefore result in a strong frustration effects which are detrimental to the convergence of our DMFT self-consistency.  To this end, for the determination of the location of magnetic instabilities, we choose another cluster which preserves the electronic structure of the triangular lattice but is in geometry more similar to the square lattice, cf.~Fig.~\ref{fig:cluster}~b).  Note that in our calculation all bonds have the same length and no additional lattice anisotropy is introduced by this geometry.  Having an even number of sites, this cluster now admits an antiferromagnetic displacement field from which we can estimate the N\'eel temperature.

\subsection{Dynamical Mean-Field Theory}

To solve the self-consistently determined impurity model we use the continuous-time quantum Monte Carlo solver in the interaction expansion (CT-INT, \cite{Rubtsov2005,Gull2011}) as implemented as an application in TRIQS \cite{TRIQS}.  With CT-INT it is possible to solve the Anderson impurity problem for large clusters. In addition this enables us to also enter the symmetry-broken phase by using a Nambu representation which doubles the number of degrees of freedom.

\subsection{Observables: Center-focused quantities}
\label{sec:observables-center-focused}

As mentioned above, based on the arguments laid out in Ref.~\cite{Klett2020}, we used the central site of the seven cluster as the closest estimate of local quantities in the thermodynamic limit.  From the physical observables of the central site we can extract a couple of figures of merit.

First we define a proxy for the spectral weight at the Fermi level which we define based on the following observation from the analytical continuation of the retarded single-particle Green function
\begin{multline}
  A(\omega \to 0) = -\frac{1}{\pi} \imag G^R(\omega \to 0) \\
  \approx -\frac{1}{\pi} \imag G_{\mathbf{R}=0}(i\omega_n \to 0) ,
\end{multline}
where we approximate the retarded Green function at zero frequency with a linear extrapolation of the impurity Green function of the central site to zero using the first two Matsubara frequencies.

The hallmark of the Mott metal to insulator transition (MIT) is a sharp drop in the spectral weight at the Fermi level as a function of interaction strength  $U/t$ which indicates the sudden formation of a gap in the single particle spectrum at $T_c^\text{MIT}$.  For $T>T_c^\text{MIT}$, however, the MIT is not a phase transition in the sense of Landau's theory of critical phenomena, and the decrease of spectral weight is spread out over a large range in parameter space.  This phenomenon is often referred to as a ``Mott crossover'' in the literature and we will adopt that terminology here as well. 

In order to define the crossover line it is identified with the inflection point of the spectral weight in parameter space, i.e., where the the derivative of the spectral weight w.r.t. the interaction strength has a minimum.

\begin{figure}[tb]
  \centering
  \unitlength=20.5pc
  \begin{picture}(1,.9)
    \put(0,.5){\includegraphics[width=\unitlength]{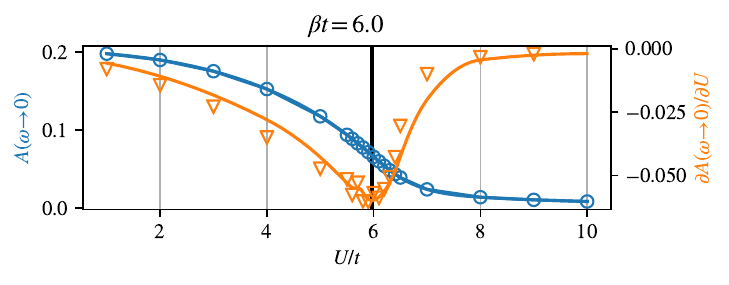}}
    \put(0,.85){a)}
    \put(0,0){\includegraphics[width=.8\unitlength]{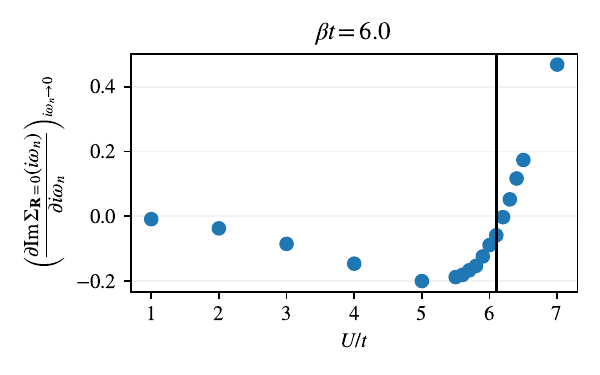}}
    \put(0,.45){b)}
  \end{picture}
  \caption{(a) Spectral weight (left $y$-axis) and its derivative (right $y$-axis) as a function of interaction. Blue circles is the spectral weight, orange downward triangles is the numerical derivative from a two-point forward finite difference stencil. The blue solid line is a B-spline fit of the spectral weight data and the orange solid line is the analytical derivative of the fitted B-spline. (b) Slope of the self-energy as a function of interaction.}
  \label{fig:center-focused}
\end{figure}

This aforementioned procedure is depicted graphically in Fig.~\ref{fig:center-focused}~a) for a fixed temperature and the interaction strength chosen as the parameter driving the transition.  The blue circles denoted our numerical data which albeit appearing smooth to the eye, does not produce a smooth derivative when taking a simple two-point forward finite difference stencil, denoted by orange triangles.  This introduces ambiguity in pinpointing the inflection point on the data.  Instead of using the numerical data directly we proceed to fit a B-spline of degree $3$ to the data using a weighted least-squares procedure to determine the coefficients by which we can also take into account the statistical error bars of our Monte Carlo data.  The first $k-1$ derivatives of a B-spline of degree $k$ are guaranteed to be smooth by construction and therefore we can obtain an unambiguous minimum of the first derivative of the fitted B-spline.  The B-spline interpolation and its first derivative are shown using solid blue and orange lines in Fig.~\ref{fig:center-focused}~a), respectively.

Second we define a proxy for a coherent quasiparticle scale.  To this end we need to specify what we mean by coherent quasiparticles.  We choose to quantify the coherence of quasiparticles by the slope of the imaginary part of the self-energy for the first two Matsubara frequencies
\begin{equation}
  \frac{\partial\imag \Sigma_{\mathbf{R}=0}(i\omega_n)}{\partial i\omega_n}\biggr|_{i\omega_n \to 0} .
\end{equation}
If this slope is positive we have a diverging self-energy which at zero temperature would indicate a gap in the single-particle spectrum.  A negative slope on the other hand indicates a finite lifetime of the quasiparticles which is why we refer to the area in the phase diagram where the slope has a negative sign as the coherent quasiparticles. In Fig.~\ref{fig:center-focused}~b) we show the evolution of the slope as a function of interaction parameter for a fixed temperature.

\subsection{Observables: Periodized quantities}
\label{sec:observables-periodized}

There exist different schemes to reperiodize impurity quantities on the cluster to the full lattice \cite{Sakai2012,Meixner2021}.  Effectively this can be seen as a restricted discrete Fourier transformation where only the modes that fit inside the cluster can be resolved.  In this work we focus on what we call the ``full periodization''.  In this case we perform the Fourier transform by summing over all combinations of cluster indices.  A complementary scheme would be the ``center-focused periodization'' in which only combinations containing the central cluster index are included in the Fourier transformation.  However, in practice, this center-focused periodization has turned out to have problems with causality in some regions of $k$-space for certain parameter regimes.

The ``full periodization'' scheme can be expressed as
\begin{equation}
  \label{eq:full-periodization}
  \mathcal{O}(\mathbf{k}) = \frac{1}{N_c} \sum_{i,j=0}^{N_c} e^{i \mathbf{k} \cdot (\mathbf{R}_i - \mathbf{R}_j)} \mathcal{O}_{i,j},
\end{equation}
where the sum runs over all $N_c$ sites in the cluster, $\mathbf{R}_i$ is the position of site $i$ in the cluster, and $\mathcal{O}_{i,j}$ is the matrix element of the observable.

\begin{figure}[tb]
  \centering
  \unitlength=20.5pc
  \begin{picture}(1,.42)
    \put(0,0){\includegraphics[width=.47\unitlength]{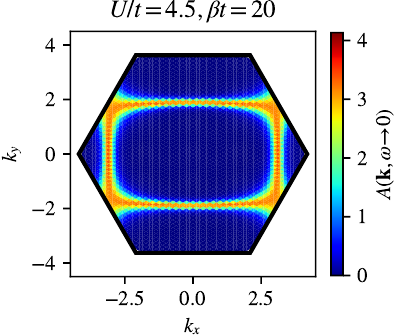}}
    \put(0,.38){a)}
    \put(.5,0){\includegraphics[width=.5\unitlength]{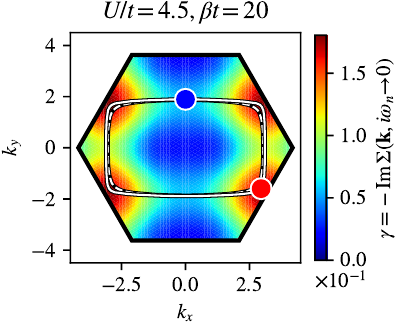}}
    \put(.5,.38){b)}
  \end{picture}
  \caption{a) The imaginary part of the periodized Green function can be used as a proxy for the spectral weight at the Fermi level which is shown by the color scale. b)~From the periodized self-energy we can extract the scattering rate as a function of momentum, plotted as the color scale.  The dashed line indicates the Fermi surface of the non-interacting model, the solid line denotes the correlated Fermi surface.  The points of smallest (largest) scattering rate on the correlated Fermi surface are identified as cold (hot) spots and indicated by a blue (red) dot.  The symmetry related partners of the cold and hot spots are omitted.}
  \label{fig:periodized}
\end{figure}

Similar to the center-focused quantities we can again define a proxy for the spectral weight at the Fermi level, but this time using the periodized Matsubara Green function $G(\mathbf{k}, i\omega_n)$
\begin{equation}
  A(\mathbf{k}, \omega \to 0) \approx - (1/\pi) \imag G(\mathbf{k}, i\omega_n \to 0) .
\end{equation}
Using the periodized impurity self-energy $\Sigma(\mathbf{k}, i\omega_n \to 0)$ we furthermore define the momentum-dependent scattering rate
\begin{equation}
  \gamma(\mathbf{k}) = - \imag \Sigma(\mathbf{k}, i\omega_n \to 0) .
\end{equation}
From the periodized self-energy we also obtain the correlated Fermi surface by solving the equation
\begin{equation}
  0 = \mu - \epsilon_{\mathbf{k}} - \real \Sigma(\mathbf{k}, i\omega_n \to 0) \iff \mathbf{k} = \mathbf{k}_F .
\end{equation}
We identify the points of smallest (largest) scattering rate on the correlated Fermi surface as ``cold'' (``hot'') spots.  In Fig.~\ref{fig:periodized}~a) we can see the spectral weight is broadened stronger around the hot spots than the cold spots indicated by colored dots in Fig.~\ref{fig:periodized}~b).  Both the spectral weight and the scattering rate indicate that we have a strong momentum differentiation of the single-particle properties, which is why we dub this system in the normal state a differentiated metal.

\begin{figure}[t!]
  \centering
  \unitlength=20.5pc
  \begin{picture}(1,1.15)
    \put(.1,.6){\includegraphics[width=.74\unitlength]{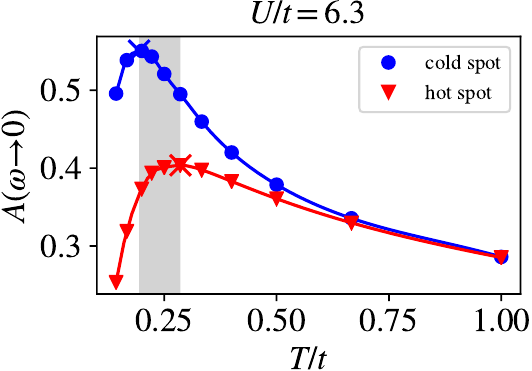}}
    \put(0,1.1){a)}
    \put(.1,0){\includegraphics[width=.8\unitlength]{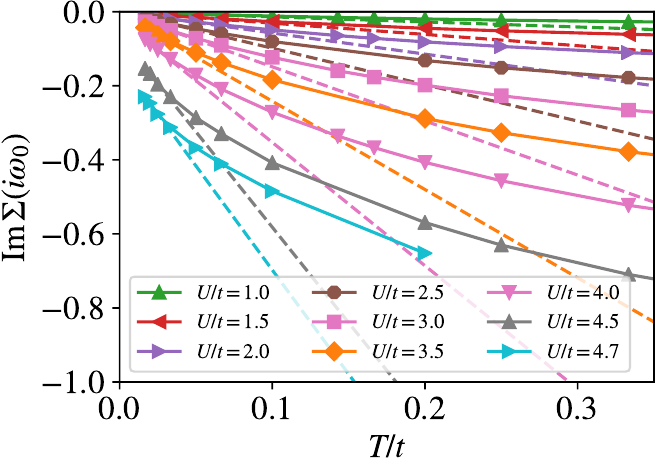}}
    \put(0,.55){b)}
  \end{picture}
  \caption{a) ``Pseudogap'' formation at the hot spot.  As a function of temperature the spectral weight at the hot spot passed through a maximum and starts decreasing before the cold spot reaches its maximum.  b) The imaginary part of the self-energy at the hot spot for the first Matsubara frequency as a function of temperature.  The dashed lines are linear fits to the lowest two temperatures.}
  \label{fig:hot-cold-spot}
\end{figure}

\begin{figure}[t!]
  \centering
  \unitlength=20.5pc
  \begin{picture}(1,1.49)
    \put(.1,1.0){\includegraphics[width=.7\unitlength]{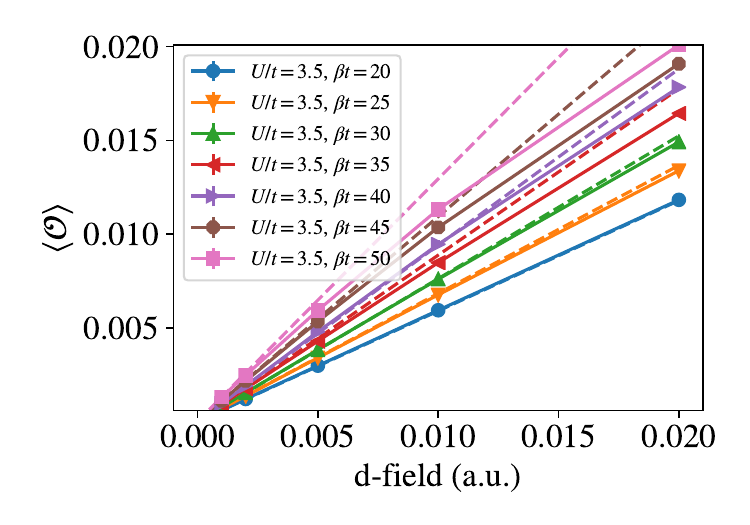}}
    \put(0,1.45){a)}
    \put(.1,.5){\includegraphics[width=.7\unitlength]{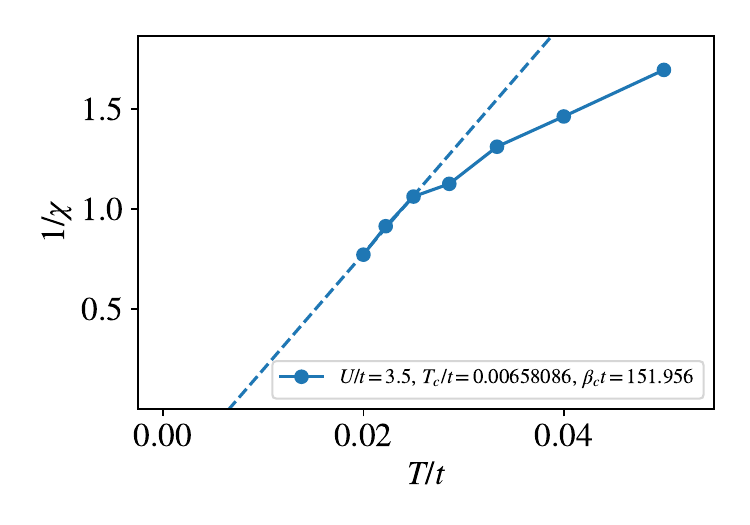}}
    \put(0,0.95){b)}
    \put(.1,0){\includegraphics[width=.7\unitlength]{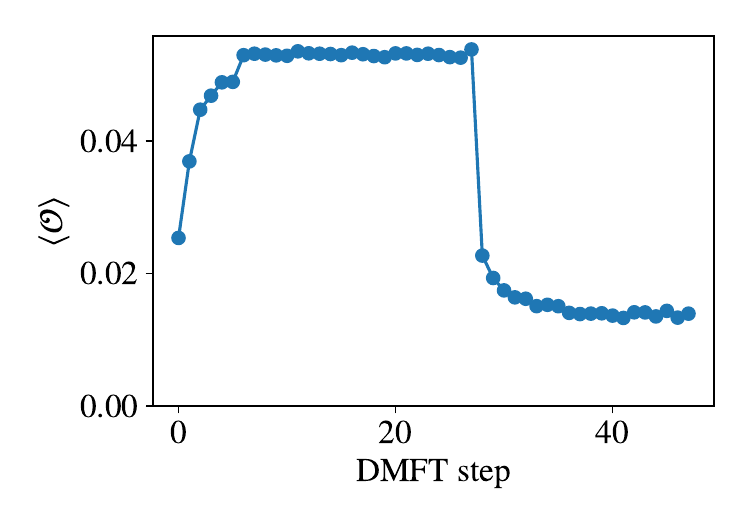}}
    \put(0,.45){c)}
  \end{picture}
  \caption{a) Expectation value of the $d$-wave superconducting order parameter as a function of the displacement field with the same symmetry.  The slope of the expectation value for small d-field is the differential susceptibility.  b) Inverse differential susceptibility as a function of temperature.  A diverging susceptibility or equivalently a zero of the inverse susceptibility signals a phase transition.  An estimate for the transition temperature is extracted by extrapolating the inverse susceptibility to zero. c) Expectation value of the superconducting order parameter below the critical temperature as a function of the DMFT step.  For the first 25 steps we apply a large displacement field which is then switched off.}
  \label{fig:linear-response}
\end{figure}

This momentum differentiation is significant since it suggests that the metallicity of the system is also momentum dependent, which naturally leads us to expect pseudogap physics.  To this end we plot the spectral weight at the hot and cold spot as a function of temperature, cf.~Fig.~\ref{fig:hot-cold-spot}~a).  We expect to see an upturn upon lowering the temperature because the spectral weight becomes sharper around the Fermi surface as thermal fluctuations are suppressed.  Close the the Mott crossover we furthermore expect that the spectral weight will first show an increase due to the aforementioned region, but then peak at a maximum and eventually decrease to zero as the Mott gap forms.  For a pseudogap we would expect the spectral weight at the hot spot to show the maximum before the cold spot upon lowering temperature.  This is what we find to some extent as shown in Fig.~\ref{fig:hot-cold-spot}~a), however, the area in parameter space where this pseudogap is present practically coincides with the Mott crossover.

In relation to superconductivity in the model it is interesting to ask what is the electronic state at low temperatures, i.e., whether it is a Fermi liquid out of which the superconducting state emerges.  Since the scattering rate is largest at the hot spot we concentrate on the self-energy at that point.  As per the ``first Matsubara frequency rule'' for a Fermi liquid the self-energy at the first Matsubara frequency must scale linearly with temperature \cite{Chubukov2012,*Maslov2012}.  In Fig.~\ref{fig:hot-cold-spot}~b) we plot the periodized self-energy at the hot spot for the first Matsubara frequency as a function of temperature and fit a linear function to the two lowest temperatures.  The temperature at which the self-energy deviates more than 30\,\% (to account for errorbars) from the fitted line is then our Fermi liquid scale.

\subsection{Ordered phases: Linear response theory}
\label{sec:ordered-phases}

To obtain phase boundaries for the different instabilities in the system we perform linear response theory.  At first a displacement field (d-field) with the same symmetry as the order parameter under investigation is applied.  Then the expectation value of the order parameter is measured and plotted against the applied d-field.  Assuming linear response the expectation should scale linearly for small values of the d-field.  Therefore we can fit a line through the origin through the first few points.  The slope of that line is the derivative of the order parameter with respect the applied field, i.e., the differential susceptibility $\chi$.  At the phase transition this susceptibility is expected to diverge by definition.  Equivalently this means that $\chi^{-1}$ goes to zero and the temperature at which it does so is the critical temperature.

In Fig.~\ref{fig:linear-response} we show the procedure on the example of the superconducting phase transition into the $d$-wave state shown in the phase diagram in the main text.  A line through the origin is fitted for small d-field for each temperature [Fig.~\ref{fig:linear-response}~a)] and the resulting susceptibility is extrapolated to obtain an estimate for the transition temperature [Fig.~\ref{fig:linear-response}~b].

For the estimation of the antiferromagnetic phase we apply a displacement field according to
\begin{equation}
  \Delta(\mathbf{R}_i) = (-1)^i \Delta_0 (c_{i,\up}^\dagger c_{i,\up} - c_{i,\dn}^\dagger c_{i,\dn}) ,
\end{equation}
in the ``artificial'' triangular geometry, cf.~Fig.~\ref{fig:cluster}~b).  Subsequently we extract the expectation value by taking the trace of the same components of the normal Green function.

\begin{figure}[t!]
    \includegraphics[width=\linewidth]{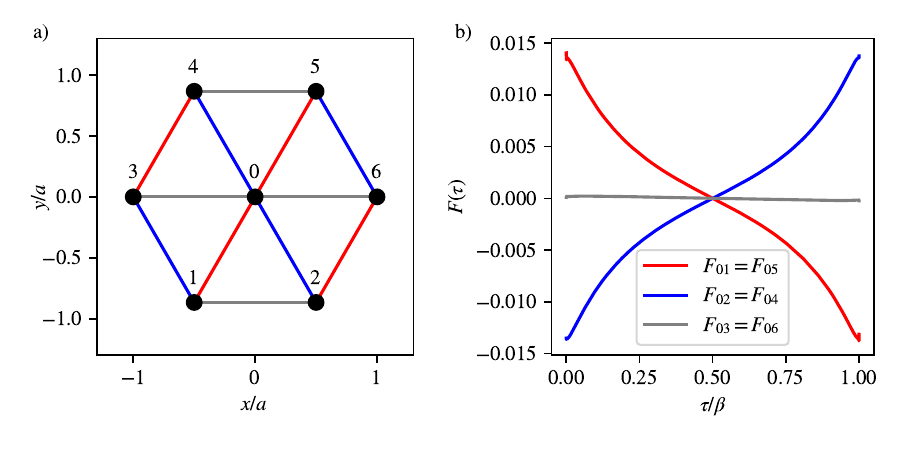}
    \caption{a) Spatial structure of the superconducting $d_{xy}$ order parameter. The bonds connecting sites $i$ to $j$ are colored by the sign of the anomalous self-energy $F_{ij}(\tau=0)$ where red~(blue) denotes positive~(negative) sign. b) Components of the anomalous Green function on the imaginary time axis for $T/t = 0.02$ and $U/t = 4.5$, i.e., within the superconducting phase of the half-filled Hubbard model on a triangular lattice with $t'/t = 0.1$.}
    \label{fig:superconductivity}
\end{figure}

\begin{figure*}[t!]
  \centering
  \includegraphics[width=\linewidth]{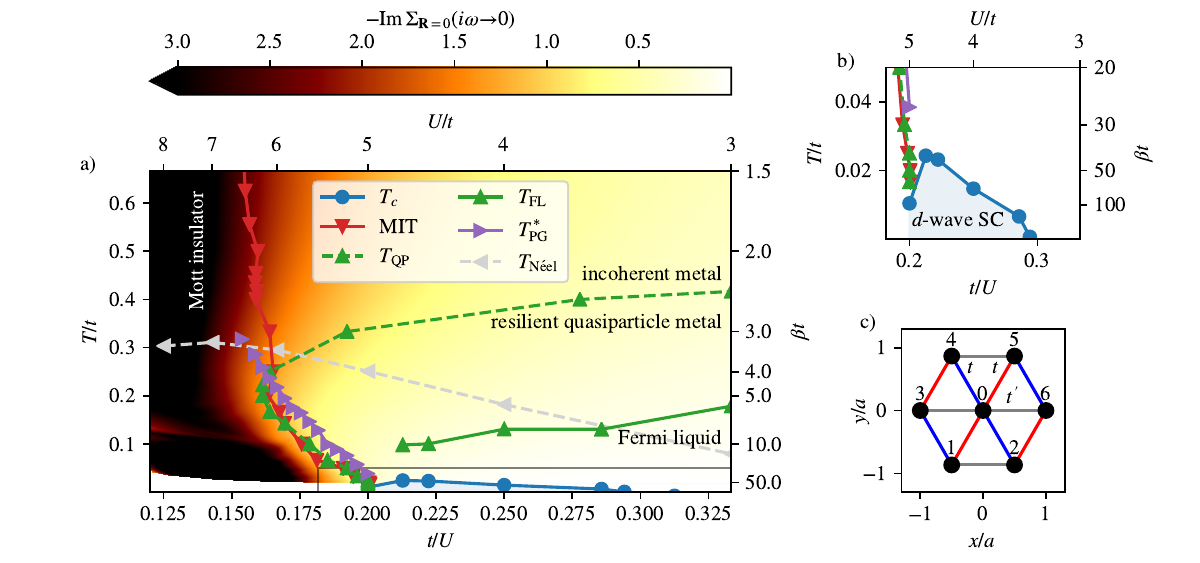}
  \caption{Analogous plot to Fig.~1 of the main text but for $t'/t = 0.1$.}
  \label{fig:phasediagram-tp0.1}
\end{figure*}

To search for superconductivity we apply a displacement field of the form~\cite{Koretsune2005}
\begin{equation}
  \begin{split}
    \Delta(\mathbf{R}_i) &= \Delta_0 \sum_l f_l (c_{i,\up} c_{i+l,\dn} + c_{i+l,\up} c_{i,\dn}) + \mathrm{h.c.} \\
    f_l &= \frac{1}{2} [
    (\delta_{l_x, 1/2} - \delta_{l_x, -1/2})
    (\delta_{l_y,\sqrt{3}/2} - \delta_{l_y,-\sqrt{3}/2})
    ]
  \end{split}
\end{equation}
in the seven site cluster, cf.~Fig.~\ref{fig:cluster}~a).  In this case we extract the traces of the appropriate components of the anomalous Green function.  For this we have to switch the entire formulation of the Hamiltonian to the Nambu basis which doubles the number of degrees of freedom.

This reformulation of the Hamiltonian in the Nambu basis allows us to directly enter the superconducting phase and perform symmetry-broken DMFT with an unbiased treatment of the anomalous part of the Green function.  To this end we first run a calculation below the critical temperature with a large displacement field applied to break the symmetry.  After the calculation has converged we switch off the displacement field and proceed to converge the now unbiased system.  If the expectation value of the order parameter converges to a finite value without the applied field we have found a superconducting phase.  The outlined procedure is shown in Fig.~\ref{fig:linear-response}~c) for $\beta t = 50$ and $U/t = 4.5$.  The resulting anomalous self-energy inside the superconducting phase is shown as a function of imaginary time in Fig.~\ref{fig:linear-response}.

\section{Magnetism and nuclear magnetic resonance}
\label{sec:magnetism}

\begin{figure}[t!]
  \centering
  \unitlength=20.5pc
  \begin{picture}(1,1.785)
    \put(.1,1.20){\includegraphics[width=.7\unitlength]{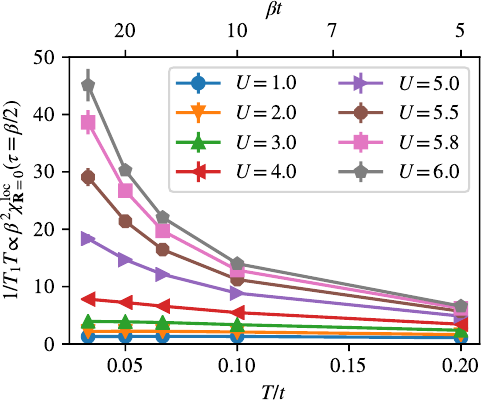}}
    \put(0,1.75){a)}
    \put(.1,.59){\includegraphics[width=.7\unitlength]{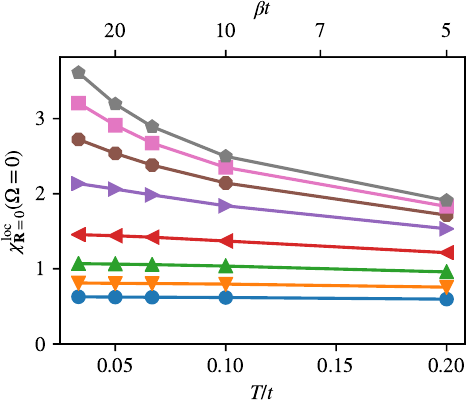}}
    \put(0,1.15){b)}
    \put(.1,0){\includegraphics[width=.7\unitlength]{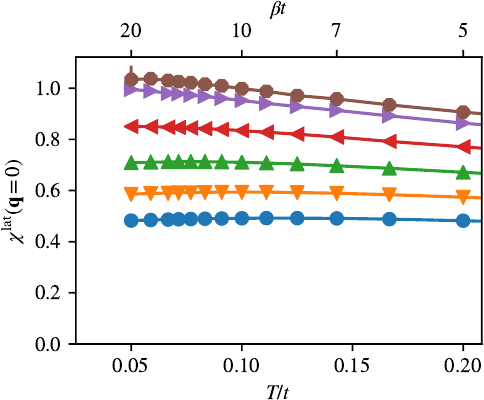}}
    \put(0,.55){c)}
  \end{picture}
  \caption{a) Spin-lattice relaxation of NMR expressed in $1/T_1T$ as a function
    of temperature and various interaction values. b) Local susceptibility of
    the impurity at the central site of the 7-site cluster as a function of
    temperature for the same interaction values.  c) Uniform susceptibility at
    $\mathbf{q} = 0$ as a function of temperature for the same interaction
    values extracted by applying a uniform external field. All panels show data
    for $t'=0.4t$ and share the same legend.}
    \label{fig:t1t}
\end{figure}

Valuable information of the spin dynamics can be obtained from nuclear magnetic resonance (NMR).
An estimate of the relaxation rate can be obtained from imaginary-time data as \cite{Randeria1992,Chen2017}:
\begin{equation}
    \frac{1}{T_1T}=\lim\limits_{\omega\rightarrow 0}\sum\limits_\mathbf{q}
    A(\mathbf{q}) \frac{\chi{''}(\mathbf{q},\omega)}{\omega}\propto \beta^2\chi_\text{loc}(\tau=\beta/2),
    \label{eq:1t1}
\end{equation}
in which we have assumed the hyperfine form factor $A(\mathbf{q})$ to be $\mathbf{q}$-independent,
a good approximation for NMR on carbon sites.
Here, $\beta=1/T$ is the inverse temperature and the local susceptibility 
\begin{equation}
    \chi_\text{loc}(\tau)=\sum_{\mathbf{q}}\chi(\mathbf{q},\tau)
\end{equation}
 is approximated by the susceptibility at the cluster's central
site $\chi^\text{loc}_{\mathbf{R}=0}$ (and the sum over $\mathbf{q}$ is normalized to the number of momentum points). The double-primed $\chi{''}(\mathbf{q},\omega)$ is shorthanded for the imaginary part of the momentum-dependent susceptibility on the real-frequency axis. Please note that by setting $\mathbf{q}=0$ one obtains the uniform susceptibility.

$1/T_1T$ is shown in Fig.~\ref{fig:t1t} a) as a function of temperature for several values of the interaction.
One can see that $1/T_1T$ is almost $T$-independent (Korringa behavior) for $t/U \gtrsim 0.25$, whereas for larger interactions the increased spin fluctuations  cause an upturn at intermediate-to-low temperatures.

Correspondingly, the local static susceptibility $\chi^\text{loc}_{\mathbf{R}=0}(\Omega\!=\!0)$ in Fig.~\ref{fig:t1t} b) increases significantly upon cooling, signalling the 
formation of local moments as the insulator is approached. 
In contrast, the uniform lattice susceptibility $\chi^\text{lat}(\mathbf{q\!=\!0},\Omega\!=\!0)$ [obtained by applying linear response theory with an applied magnetic field, analogously to Sec.~\ref{sec:ordered-phases}] increases at a much smaller rate in Fig.~\ref{fig:t1t} c), as expected since 
it is sensitive to the antiferromagnetic superexchange interaction. 
Note however that our cluster has an odd number of sites and hence the isolated cluster has 
a spin-1/2 ground-state, favoring local moment physics in the insulator. Hence, definitive conclusions 
about magnetism must be drawn with care and presumably require a comparison to clusters 
with even sites as well \cite{HebertMSc}. 
Nonetheless, the increase upon cooling of $1/T_1T$ at higher $T$ is in reasonable agreement with experiments. 
We also note that due to the finite cluster size our calculations are most likely oblivious to long-range superconducting fluctuations which may cause a downturn of $1/T_1T$ at low $T$ \cite{Furukawa2023,Imajo2023}.

\section{Results for lower frustration}
\label{sec:additional}

In the main text we have presented a comprehensive phase diagram of the triangular lattice Hubbard model for a hopping anisotropy of $t'/t = 0.4$.  Here we also present data for a smaller value of the anisotropy  $t'/t = 0.1$ and compare and contrast the findings with the ones presented in the main text.

In Fig.~\ref{fig:phasediagram-tp0.1} we show the same phase diagram as Fig.~1 in the main text, but for $t'/t = 0.1$.  We find an overall qualitative similarity between the two albeit with two important differences:  First of all the smaller anisotropy brings the system closer to the geometry of the square lattice and therefore we find a much larger antiferromagnetic phase.  At the same time we note that the other phase transitions and crossovers (MIT, superconductivity, ``pseudogap'') move to smaller $U/t$ or equivalently larger $t/U$.

\begin{figure}[t!]
    \includegraphics[width=\linewidth]{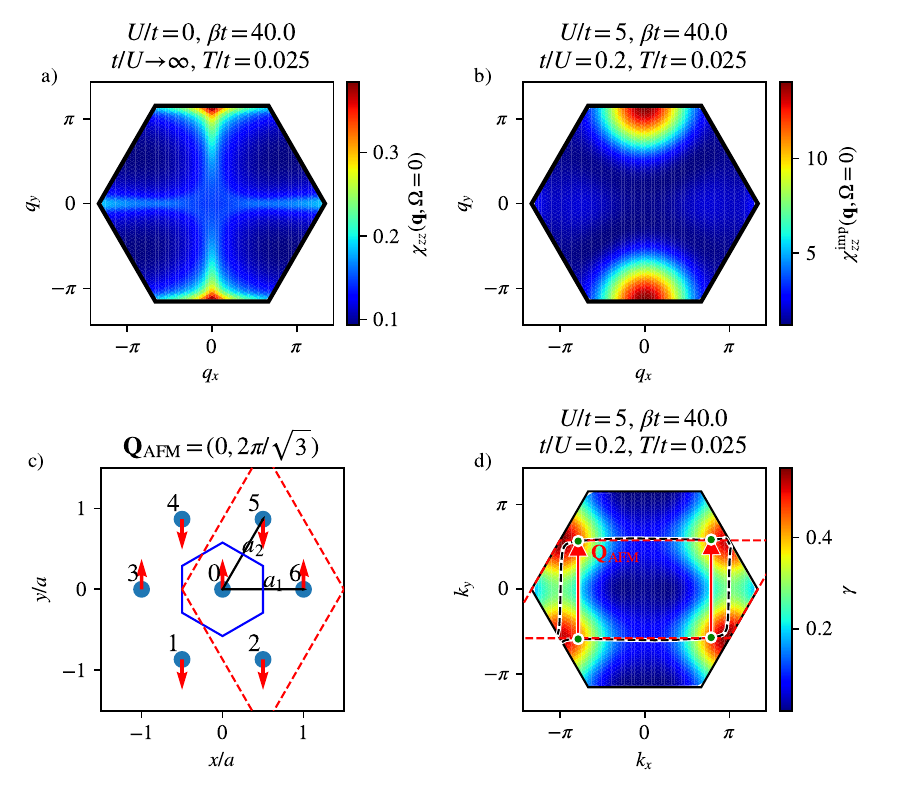}
    \caption{Analogous plot to Fig.~3 of the main text but for $t'/t = 0.1$.}
    \label{fig:nesting-tp0.1}
\end{figure}

\begin{figure*}[tb!]
  \centering
  \includegraphics[width=\linewidth]{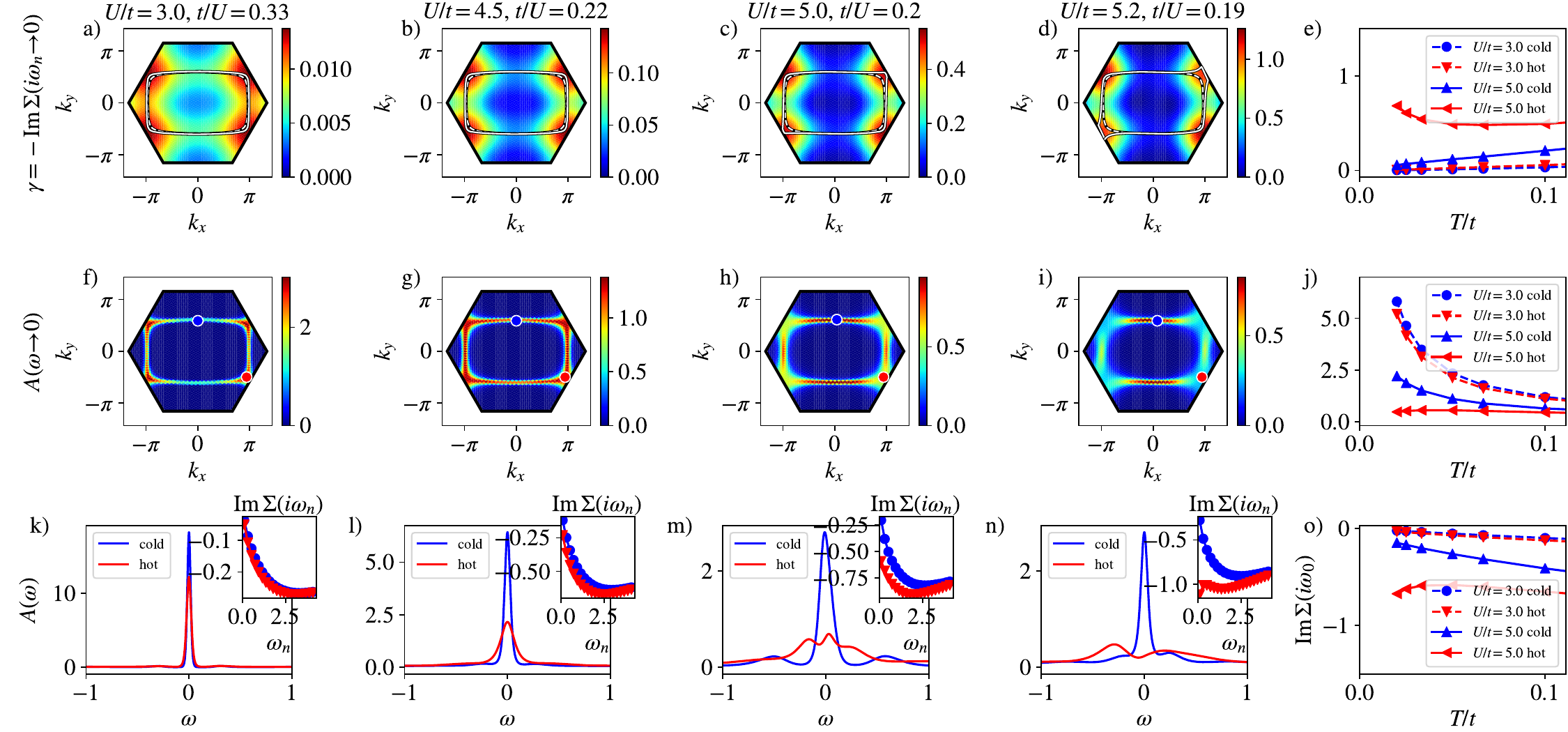}
  \caption{Analogous plot to Fig.~2 of the main text but for $t'/t = 0.1$.}
  \label{fig:periodization-tp0.1}
\end{figure*}

The various lines in the phase diagram are extracted using the techniques detailed in the previous sections.  From the single-particle center-focused observables described in Sec.~\ref{sec:observables-center-focused} we extract the MIT and the quasiparticle coherence scale $T_{\mathrm{QP}}$.  Furthermore we plot the imaginary part of the self-energy at the central site as a color scale in the background.  From the periodized observables described in Sec.~\ref{sec:observables-center-focused} we extract the ``pseudogap'' transition $T_{\mathrm{PG}}$ and the Fermi liquid temperature $T_{\mathrm{FL}}$.

Turning to the nature of the magnetic phase we show in Fig.~\ref{fig:nesting-tp0.1}a) the non-interacting static spin-spin susceptibility $\chi_{zz}(\mathbf{q},i\Omega_n=0)$ (Lindhard bubble) for $t'/t=0.1$.  In contrast to the Lindhard bubble for $t'/t=0.4$ in the main text, we do not find a splitting of the peak at $\mathbf{q}=\mathbf{M}=(\pi,0)$. As for $t'/t=0.4$ this corresponds to stripy antiferromagnetism with ferromagnetic correlations along the bonds with suppressed hopping $t'$, cf.~Fig.~\ref{fig:nesting-tp0.1}c). We also show the interacting susceptibility in momentum space $\chi_{zz}(\mathbf{q},i\Omega_n=0)$ at $t/U=0.2$ in Fig.~\ref{fig:nesting-tp0.1}b) obtained from the cluster susceptibility by the full periodization scheme described in Sec.~\ref{sec:observables-periodized}. The periodized $\chi_{zz}$ is also peaked at $\mathbf{M}$.

Finally we investigate the nature of the metallic phase by looking at the evolution of various Fermi surface observables as a function of interaction in Fig.~\ref{fig:periodization-tp0.1}.  To this end we periodize the self-energy according to Sec.~\ref{sec:observables-periodized} and extract the various observables described in that section. The first row of plots a)--d) shows the scattering rate and we can see that for larger interaction the scattering rate picks up in magnitude and becomes more differentiated throughout the Brillouin zone leading to the formation of regions of large (hot spots) and small (cold spots) scattering rate. This is accompanied by a deformation of the Fermi surface and goes hand in hand with a reduction of the spectral weight at the hot spots shown in the second row f)--i). The loss of spectral weight at the hot spot is reminiscent of a pseudogap phase. To this end we show in the right column in e) the scattering rate, in j) the spectral weight, and in o) the self-energy at the first Matsubara frequency at the hot and cold spot as a function of temperature for two different values of $U/t$. For $U/t=3.0$ we find Fermi liquid-like behavior, whereas for $U/t=5.0$ we find metallic behavior at high temperature which becomes a bad metal at low temperatures with increasing scattering rate, decreasing spectral weight, and an inflection in the self-energy for the first Matsubara frequency at the hot spot.

\FloatBarrier
\bibliographystyle{unsrturl}
\bibliography{references.bib}